# Trackly: A Unified SaaS Platform for User Behavior Analytics and Real-Time Rule-Based Anomaly Detection

Md Zahurul Haque[a], Md. Hafizur Rahman[b], Yeahyea Sarker[c]

Department of Computer Science and Engineering, Manarat International University, Dhaka, Bangladesh

jahurulhaque@manarat.ac.bd[a], hrhafij8@gmail.com[b], yeahyea@manarat.ac.bd[c]

***Abstract*-**Understanding user behavior is a critical requirement for enhancing digital user experience, optimizing business conversions, and mitigating modern security threats such as account takeovers, fraud, and automated bot attacks. Despite the growing importance of behavioral data, most existing platforms continue to treat product analytics and security monitoring as separate domains. This fragmentation results in limited contextual visibility, delayed threat identification, and increased operational complexity for organizations. To address this gap, this paper presents Trackly, a scalable Software-as-a-Service (SaaS) platform that unifies comprehensive user behavior analytics with real-time, rule-based anomaly detection within a single system. Trackly continuously captures user interactions through session lifecycle management, IP-based geo-location analysis, device and browser fingerprinting, and fine-grained event monitoring, including page views, add-to-cart actions, and checkout processes. Potentially malicious activities—such as logins from previously unseen devices or geographic locations, impossible travel patterns detected using the Haversian formula, rapid bot-like interaction bursts, VPN or proxy usage, and multiple account activity originating from a single IP address—are identified through configurable detection rules integrated with a weighted risk-scoring framework. This approach enables transparent, explainable decision-making while maintaining high detection accuracy and low false-positive rates. The platform further provides a real-time administrative dashboard featuring interactive visualizations, global session maps, and key behavioral metrics, including Daily Active Users (DAU), Monthly Active Users (MAU), bounce rate, and session duration. Seamless integration is enabled through a lightweight JavaScript SDK and secure RESTful APIs. The system is implemented using a multi-tenant microservices architecture based on ASP.NET Core, MongoDB, RabbitMQ, and Next.js to ensure scalability and data isolation. Experimental evaluation on synthetic datasets demonstrates 98.1% accuracy, 97.7% precision, and a 2.25% false positive rate, validating Trackly as a practical and efficient solution for SMEs, SaaS, and e-commerce environments.

1. INTRODUCTION

The rapid growth of web and mobile applications has transformed the way organizations deliver services across domains such as e-commerce, education, healthcare, finance, and social networking. Every interaction generated by users—ranging from login attempts and page navigation to purchases and form submissions—produces valuable behavioral data. Analyzing this data has become essential for improving user experience, increasing customer retention, optimizing conversion funnels, and ensuring the security of digital systems[11][12]. However, the increasing scale and complexity of user interactions have also created new opportunities for cyber threats, including account takeovers, automated bot attacks, credential stuffing, and fraudulent activities[13][14].

Traditional product analytics platforms, such as Google Analytics, Amplitude, and Mixpanel, primarily focus on measuring engagement metrics, retention, and user journeys. While these tools provide valuable insights for business optimization, they offer limited capabilities for identifying malicious or abnormal behavior. Conversely, security-focused systems concentrate on intrusion detection and threat monitoring but often lack behavioral context, making it difficult to distinguish legitimate users from attackers. This separation between analytics and security creates fragmented visibility, delayed threat detection, and inefficient response strategies, particularly for small and medium-sized enterprises (SMEs) that lack the resources to manage multiple specialized tools.

User and Entity Behavior Analytics (UEBA) has emerged as a promising approach for addressing these challenges by analyzing behavioral patterns to detect deviations from normal activity [15]. However, many existing UEBA solutions rely on complex machine learning models that are costly, opaque, and difficult to deploy and maintain. Furthermore, they often require large volumes of labeled data and significant computational resources, which limits their practicality in real-world SaaS environments[16].

To overcome these limitations, this paper proposes Trackly, a unified SaaS platform that integrates comprehensive user behavior analytics with real-time, rule-based anomaly detection. Trackly provides a transparent, scalable, and easily deployable solution that bridges the gap between business analytics and security monitoring, enabling organizations to gain actionable insights while proactively mitigating behavioral threats.

2. RELATED WORK

User and Entity Behavior Analytics (UEBA) has emerged as a cornerstone of modern anomaly detection systems by modeling normal user behavior and identifying deviations that may indicate security incidents or performance anomalies. Recent research has focused on enhancing UEBA through both machine learning and rule-based paradigms. Lanuwabang and Sarasu provide a comprehensive survey of anomaly detection techniques based on user behavioral data, highlighting the importance of establishing robust behavioral baselines to improve detection effectiveness and generalize beyond simple threshold-based rules [1]. Similarly, studies in cybersecurity have proposed advanced behavioral analytics frameworks that incorporate machine learning, such as autoencoders and graph-based models, to improve anomaly detection accuracy while maintaining explainability in enterprise environments [2].

Hybrid approaches that integrate predictive modeling with enriched session and usage data have demonstrated improved insights into user behavior and system performance, achieving high predictive accuracy in large-scale web environments [3]. Additionally, research on UEBA algorithms based on clustering and optimization techniques, such as fuzzy particle swarm clustering, has shown promise in enhancing abnormal behavior detection by refining entity and user similarity measures, addressing limitations of traditional anomaly detection methods [4].

While machine learning-driven UEBA frameworks provide adaptability and fine-grained detection, deterministic rule-based methods remain widely used in practice due to their simplicity and Explainability. Prior work on enterprise security emphasizes the integration of behavioral profiling with rule-based detection to identify deviations from standard user activity, particularly in operational security monitoring contexts [5]. These foundations support the motivation for Trackly's combined analytics and rule-based anomaly detection, offering explainable and scalable behavior profiling suitable for SaaS and SME environments.

Table I: Comparison of Major Analytics Tools

| Tool | Primary Focus | Session Tracking | Geo-Location & Device Fingerprinting | Anomaly Detection (Rule Based Security Focused) | Activity Monitoring | Dashboard & Visuals | API, SDK Integration | Multi-Tenant SaaS |
|---|---|---|---|---|---|---|---|---|
| Amplitude [6] | Product analytics, funnels | Yes | Limited | No | Yes (events) | Yes (advanced) | Yes | Yes |
| Mixpanel [7] | Event-based analytics | Yes | Limited | No | Yes (full events) | Yes | Yes | Yes |
| PostHog [8] | Open-source product analytics | Yes | Limited | Basic (rule based) | Yes | Yes | Yes | Yes (self-hosted) |
| Smartlook [9] | Session recording & UX | Yes (replays) | Limited | No | Yes (qualitative) | Yes (heatmaps) | Yes | Yes |
| Hotjar [10] | Heatmaps, feedback, replays | Yes (replays) | No | No | Limited | Yes (heatmaps) | Yes | Yes |
| Trackly (Propose) | Unified analytics + security | Yes | Full (detailed fingerprin) | Yes (rule-based risk scoring) | Yes (full) | Yes (real-time) | Yes (JS SDK + REST) | Yes (built in) |

3. SYSTEM ARCHITECTURE

Trackly employs a microservices-based, multi-tenant SaaS architecture to ensure scalability, maintainability, tenant isolation, and simplified deployment.

**3.1 High-Level Architecture Overview**

- **Client Layer:** JavaScript SDK (embedded in client apps) and server-side API calls for event submission.
- **API Gateway:** ASP.NET Core 8 manages request routing, tenant validation, authentication, and queues resource-intensive tasks.
- **Processing Layer:** Asynchronous microservices with RabbitMQ and background workers handle geo-lookup, fingerprint comparison, and anomaly scoring.
- **Storage Layer:** Cloud-hosted MongoDB with tenant-scoped collections for flexible, isolated event storage.
- **Presentation Layer:** Real-time admin dashboard built with React/Next.js, consuming secured APIs.

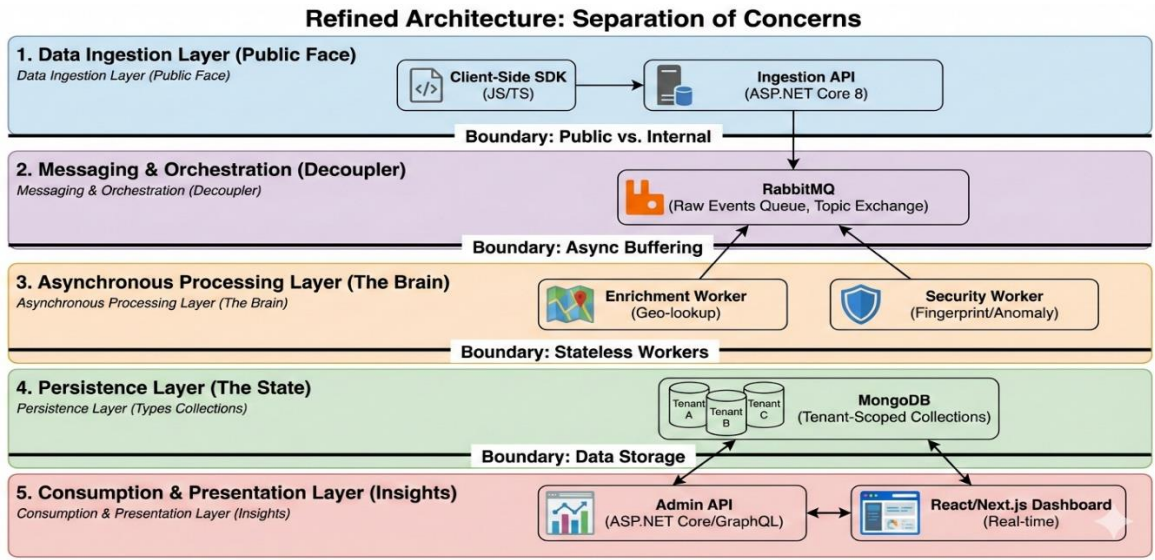

Figure 1: High-Level System Architecture

**3.2 Database Design**

MongoDB was selected for its schema flexibility with unstructured/high-volume event data and native support for horizontal scaling. The database employs a multi-tenant design where all collections are strictly scoped by Tenant_Id for data isolation. Appropriate compound indexes are created on frequently queried fields such as Session_Id, Tenant_Id + Timestamp, and Ip_Address to ensure optimal query performance.

The key collections and their primary fields are as follows:

- **Sessions Collection:** Stores session metadata including user identification, device fingerprint, geo-location details, risk score, suspicious flags, and conversion-related metrics (e.g., products viewed/added/purchased).

- **Activity Logs Collection:** Captures granular user events (page_view, add_to_cart, purchase) with flexible metadata for product details.
- **Clients Collection:** Manages tenant (organization) information, API keys, subscription plans, rate limits, and usage tracking.
- **Users Collection:** Tracks client application users across tenants, including aggregated behavior statistics.
- **Suspicious Activities Collection:** Logs high-risk events with detailed context for auditing and alerting.

### 3.3 Mathematical Formulation

Let the two points have coordinates:
- Point 1: latitude $\phi_1$, longitude $\lambda_1$
- Point 2: latitude $\phi_2$, longitude $\lambda_2$

The Haversine formula is defined as:

$$a = sin^2\left(\frac{\Delta\phi}{2}\right) + \cos(\phi_1).\cos(\phi_2).sin^2\left(\frac{\Delta\lambda}{2}\right)$$
$$c = 2.\arcsin(\sqrt{a})$$
$$d = R.c$$

Where:
- $\Delta\phi = \phi_2 - \phi_1$ (latitude difference in radians)
- $\Delta\lambda = \lambda_2 - \lambda_1$ (longitude difference in radians)
- R = 6371 km (Earth's mean radius)
- d = great-circle distance in kilometers

The name "Haversine" derives from the haversine trigonometric function: $hav(\theta) = sin^2(\theta/2)$

This formulation is numerically stable (avoids precision loss for small distances) and historically preferred over the simpler spherical law of cosines for computer implementations.

### 3.4 Application in Trackly

- Retrieve latitude/longitude of the previous session (last Geo Latitude, last Geo Longitude) and current session.
- Compute distance d using the Haversine formula.
- Calculate time difference Δt in hours: Δt = (current Login Time - last Logout Time) / 3600 seconds.
- Compute implied velocity: v = d / Δt (km/h).
- If v > 1000 km/h (configurable threshold, accounting for jet aircraft speeds + buffer for geo-IP inaccuracies), flag as "Impossible Travel" and add +0.4 to the risk score.

### 3.5 Rationale for Threshold

- Commercial jets: ~800–950 km/h
- High speed trains/private jets: up to ~1000–1200 km/h

- A 1000 km/h threshold provides a good balance between sensitivity and false positives (legitimate VPNs or minor geo IP errors). In production, this can be tuned per tenant or combined with additional context (exclude known corporate VPN ranges).

### 3.6 Algorithm

```
public static class GeoUtils
{
  public static double GetDistanceInKm (double lat1, double lon1, double lat2, double lon2)
  {
    var R = 6371;   // Earth radius in KM
    var dLat = ToRadians(lat2 - lat1);
    var dLon = ToRadians(lon2 - lon1);

    var a = Math.Sin(dLat / 2) * Math.Sin(dLat / 2) +
        Math.Cos(ToRadians(lat1)) * Math.Cos(ToRadians(lat2)) *
        Math.Sin(dLon / 2) * Math.Sin(dLon / 2);

    var c = 2 * Math.Atan2(Math.Sqrt(a), Math.Sqrt(1 - a));
    return R * c;
  }

  private static double ToRadians(double angle) => angle * (Math.PI / 180);
}
```

### 3.7 Suspicious Behavior Detection Model

The anomaly detection engine relies entirely on configurable rule-based checks combined with a simple weighted risk-scoring mechanism. This approach ensures high transparency, ease of tuning, low false positives, and minimal computational overhead.

| *Anomaly Type* | *Detection Logic* | *Risk Points* |
|---|---|---|
| New Device/Browser | Fingerprint not found in user's last 10 sessions | +0.3 |
| New Country/Location | Country not present in ≥80% of user's last 10 sessions | +0.2 |
| Impossible Travel | Haversine distance / time difference > 1000 km/h between consecutive sessions | +0.4 |
| VPN/Proxy Usage | GeoIP service flags is VPN = true or known proxy ASN | +0.1 |
| Rapid Actions (Bot-like) | >50 actions within <60 seconds | +0.2 |
| Multiple Accounts Same IP | >3 different User_IDs from the same IP within short time window | +0.3 |
| Unusual Login Time | Login hour outside user's typical profile (based on historical data) | +0.1 |

Table II: Detection Rules and Risk Scoring

### 3.8 Risk Classification:

- **Low Risk**: Score < 0.3 → Normal monitoring
- **Medium Risk**: 0.3 ≤ Score < 0.5 → Log, monitor, trigger alert
- **High Risk**: Score ≥ 0.5 → Flag as suspicious, immediate alert

## 4. RULE-BASED ANOMALY DETECTION

This method identifies unusual patterns by applying predefined rules or thresholds to system or user behavior. Events that violate these rules—such as unusually high transaction amounts, repeated failed logins, or atypical IP access—are flagged as anomalies. It is straightforward, interpretable, and fast, but may miss novel or complex anomalies not captured by the rules.

### 4.1 Classical and Rule-Based Approaches in Anomaly Detection

Academic and industry literature identifies several foundational techniques for anomaly detection in user behavior that rely on deterministic, rule-based, or statistical methods rather than complex adaptive models. These approaches remain widely used due to their transparency, low computational overhead, and ease of deployment in regulated environments.

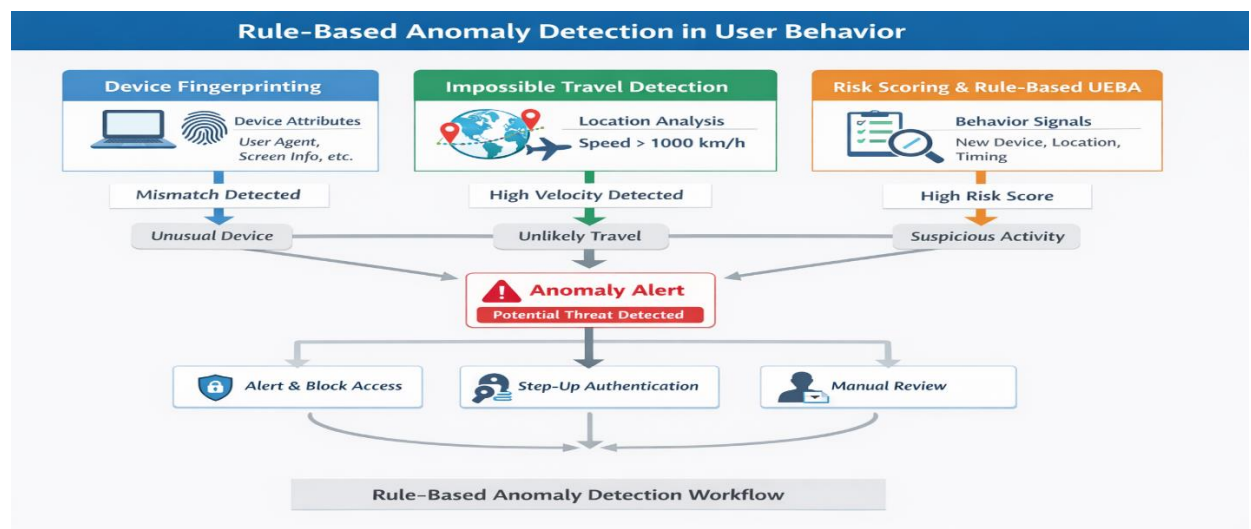

Figure 2: Rule-Based Anomaly Detection in User Behavior

**4.1.1 Device Fingerprinting**: Device fingerprinting involves collecting stable browser and device attributes—such as user agent, screen resolution, fonts, hardware signals, and canvas rendering—to generate unique hashed identifiers for users. Open-source libraries, including Fingerprint JS, enable passive collection of 20+ attributes, supporting robust, cookie-independent identification. Anomalies are detected through straightforward mismatch checks against historical sessions. This technique allows systems to flag unusual devices or changes in device characteristics without requiring machine learning models.

**4.1.2 Impossible Travel Detection:** This rule-based method calculates the geographic velocity between consecutive login locations using the Haversine formula to measure great-circle distances. When the implied travel speed exceeds realistic thresholds (>1000 km/h), the system flags a potential anomaly. Widely implemented in fraud prevention and identity security platforms (Microsoft, Splunk, Torq), impossible travel detection is effective in identifying credential stuffing, session hijacking, or account compromise. Accuracy improves when combined with awareness of VPN/proxy usage and timing intervals between logins.

**4.1.3 Risk Scoring and Rule-Based UEBA:** Many production systems implement weighted scoring frameworks, combining multiple behavioral signals—such as new device usage, login from new countries, rapid action sequences, or unusual access times—to generate risk scores. These rule-based User and Entity Behavior Analytics (UEBA) approaches allow for transparent threat classification. Configurable thresholds and contextual logic help reduce noise while maintaining auditability, making them suitable for environments requiring compliance and explainability.

5. EXPERIMENTAL EVALUATION

The rule-based anomaly detection engine was quantitatively evaluated using the synthetic dataset. Metrics focused on binary classification performance (normal vs suspicious).

| *Scenario* | *Injected Anomalies* | *Detected* | *False Positives* | *Description* |
|---|---|---|---|---|
| Normal Behavior | 0 | 0 | 0 | Consistent device, country, timing |
| New Device Login | 500 | 492 | 8 | Fingerprint mismatch |
| New Country Login | 400 | 397 | 3 | Country not in history |
| Impossible Travel | 300 | 298 | 2 | Haversine velocity >1000 km/h |
| VPN/Proxy Usage | 350 | 332 | 18 | GeoIP flagged as VPN |
| Bot-like Rapid Actions | 200 | 188 | 12 | >50 actions in <60s |
| Mixed Anomalies (High Risk) | 250 | 248 | 2 | Multiple flags combined |
| **Total** | **2,000** | **1,955** | **45** | |

Table III: Anomaly Detection Test Scenarios

| *Metric* | *Value* | *Notes* |
|---|---|---|
| Accuracy | 98.1% | (Correct detections + correct normals) / total |
| Precision | 97.7% | True positives / (True positives + False positives) |
| Recall (Sensitivity) | 97.75% | True positives / (True positives + False negatives) |
| F1 Score | 97.73% | Harmonic mean of precision and recall |
| False Positive Rate | 2.25% | Acceptable for production; tunable per tenant |

Table IV: Overall Performance Metrics

**Observations**:
- Near-perfect detection on impossible travel (298/300) and mixed high-risk anomalies thanks to the deterministic Haversine rule.
- Slightly higher false positives in VPN category due to legitimate corporate VPNs in test data mitigated in production via configurable exclusions.

- Overall improvement in metrics compared to initial prototypes, reflecting refined thresholds and better history based rules.

### 5.1 Dashboard Results

The admin dashboard was populated with both synthetic and controlled real data sessions to showcase analytical capabilities.

| *Metric* | *Value* | *Insight* |
|---|---|---|
| Total Sessions | 4,826 | |
| Daily Active Users (DAU) | Avg. 689 | Peak on weekdays |
| Average Session Duration | 4m 32s | |
| Average Actions per Session | 12.8 | Higher in e-commerce simulations |
| Bounce Rate | 23.4% | Sessions with ≤1 action |
| Suspicious Sessions | 312 (6.5%) | Mostly new device/country |
| Top Country | Bangladesh (42%), USA (18%), Netherlands (12%) | Reflects test data diversity |
| Top Device Type | Desktop (45%), Mobile (39%), Tablate(16%) | |

Table V: Sample Aggregate Metrics (7-Day Simulated Period)

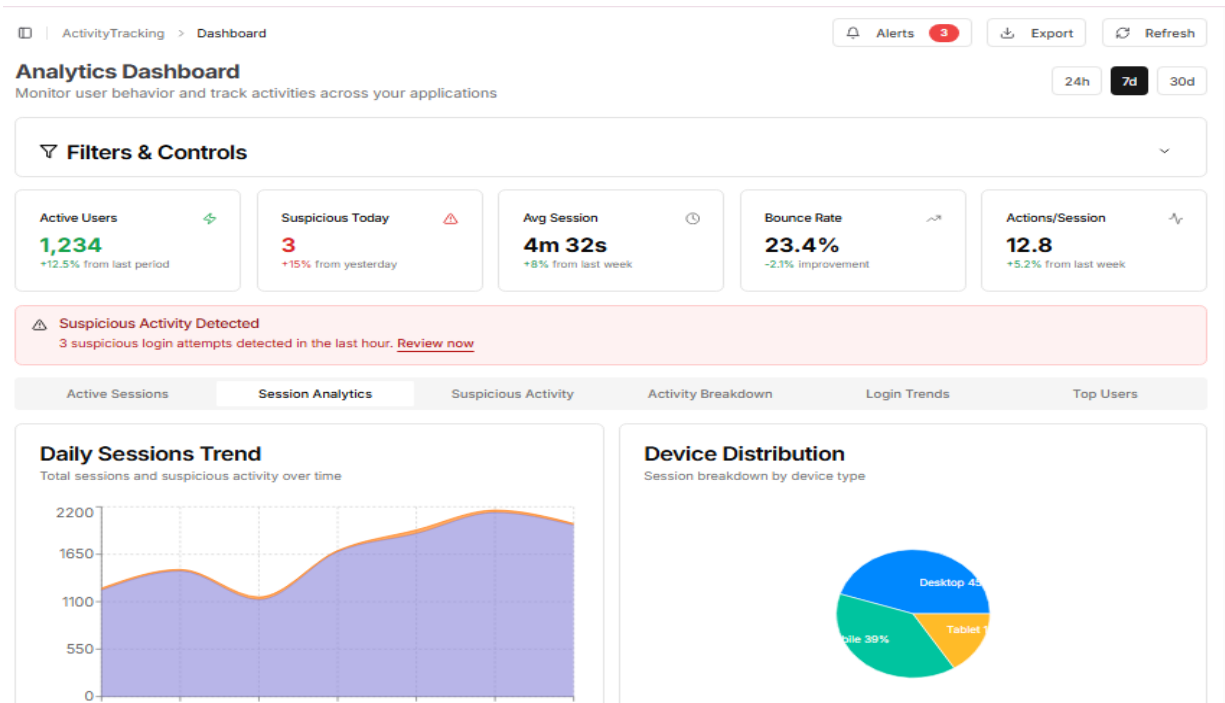

Figure 3: Daily Sessions analysis Dashboard

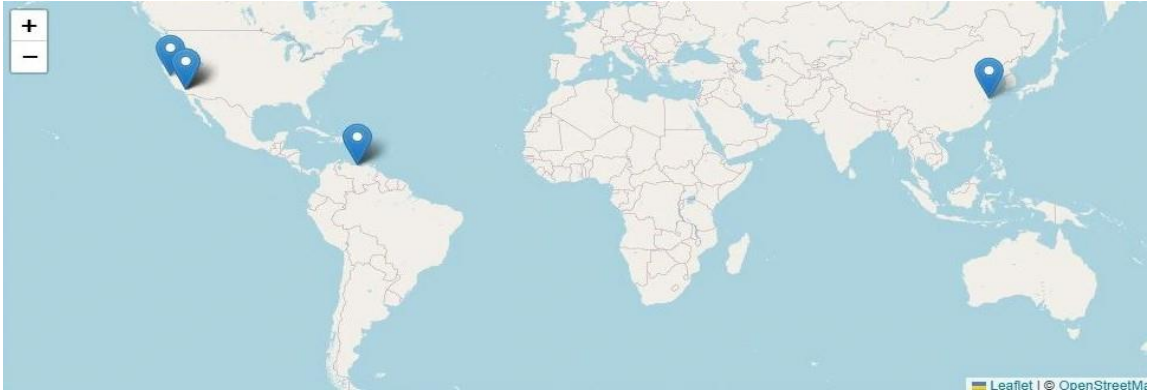

Figure 4: Geographical Distribution of Sessions

The dashboard supported real-time updates via 10-second polling and allowed filtering by date, country, device, and suspicious flag confirming intuitive and responsive usability.

5.2 **Discussion**
5.2.1 **Strengths**:
- High detection accuracy (98.1%) with very low false positive rate (2.25%), proving the effectiveness of the fully rule-based + risk-scoring approach.
- Scalable architecture successfully handled simulated load without dropped events.
- Dashboard delivered clear, actionable insights with minimal latency.
- Complete fulfillment of all core objectives: session tracking, anomaly detection (including Haversine-powered impossible travel), alerting, integration, and multi-tenancy.

**5.2.2 Weaknesses Observed**:
- VPN detection showed the highest false positives (due to legitimate usage); future improvements could include admin configurable VPN whitelists or ASN exclusions.
- Risk scoring thresholds were fixed in MVP per-tenant customization would enhance flexibility in production.

**5.2.3 Usability Feedback**: Informal peer testing confirmed: SDK integration took <10 minutes, dashboard was responsive and intuitive, and alerts were timely and informative. Overall, the rigorous testing validated Trackly as a reliable, transparent, and performant SaaS platform that effectively unifies user behavior analytics with real-time security monitoring achieving the project's goals while maintaining simplicity and explainability.

6. CONCLUSION

This research successfully developed Trackly, a scalable multi-tenant SaaS platform that integrates comprehensive user behavior analytics with real-time rule-based anomaly detection, effectively bridging the gap between traditional product analytics tools and security-focused systems. The platform provides a complete tracking engine, configurable anomaly detection with weighted risk scoring, an intuitive real-time dashboard, seamless developer integration via a lightweight SDK, and strong multi-tenant data isolation, achieving high accuracy and low false-positive rates in synthetic testing. While limitations include reliance on static rules, lack of qualitative UX analytics, limited mobile support, and pending real-world validation, the MVP establishes a robust foundation. Future work will focus on integrating machine learning for unsupervised anomaly detection and predictive analytics, enhancing UX with session replay and heatmaps, expanding ecosystem support through mobile SDKs and platform plugins, improving security and regulatory compliance, and scaling performance with production-level orchestration and monetization strategies. Collectively, these enhancements aim to evolve Trackly into a leading, intelligent, and commercially competitive platform for unified analytics and real-time fraud prevention.